\documentstyle[preprint,aps]{revtex}
\begin{document}
\begin{titlepage}
\begin{centering}
\title{
${\cal O}(\alpha_{s})$ Spin-Dependent Weak Structure Functions
\thanks{Partially supported by CONICET-Argentina.}}
\author{D. de Florian and R.Sassot$^{1}$ }
\address{
Laboratorio de F\'{\i}sica Te\'{o}rica \\
Departamento de F\'{\i}sica \\
Universidad Nacional de La Plata \\ C.C. 67 - 1900 La Plata \\
Argentina }
\address{
$^{1}$Departamento de F\'{\i}sica \\
Universidad de Buenos Aires \\
Ciudad Universitaria, Pab.1 \\
(1428) Bs.As. \\
Argentina\\}
\maketitle
\end{centering}
\centerline{ La Plata Preprint 11-94 \ \ \ (6 December 1994)}
\begin{abstract}
The complete next to leading logarithmic (${\cal O}(\alpha_{s})$)
corrections to the spin dependent weak deep inelastic structure functions
$g_{1},g_{3}$ and $g_{4}$, are calculated using dimensional regularization
within the HVBM
method. Analysing the quark and gluon initiated contributions to these
corrections for different values for the quark masses, a consistent
factorization prescription for spin dependent quark distributions, which
safely removes soft contributions, is defined.
It is shown that within this scheme, quark initiated corrections are comparable
in magnitude to those of gluonic origin, even though their contributions to the
moments are small.
\end{abstract}

\end{titlepage}
\noindent {\large \bf Introduction}\\

The very recent release of new and more precise data on polarized deep
inelastic scattering experiments \cite{exp} have considerably increased the
interest
in many aspects of the internal spin structure of nucleons and their related
experiments \cite{forte}. Among them, a growing number of parametrizations for
polarized
parton distributions, designed to reproduce the experimental data, and cross
sections for spin dependent processes, thought as further constraints for the
distributions and cross check for the data, are available at present
\cite{dist,dist2,dist3}.

However, the increasing precision of the experiments and the alternative of
having large contributions from gluon initiated processes, require both parton
distributions and cross sections being obtained at least at order $\alpha_{S}$
and with a consistent prescription for the re\-gu\-la\-rization and
substraction of
soft contributions. It has been shown \cite{vogel}, in the context of the
photon-gluon fusion
process, that a careful analysis of the regularization procedures for collinear
singularities is essential for isolating  and substracting soft contributions.
This ana\-ly\-sis not only sheds light on the controversy about this point, but
emphasizes the importance of working in a consistent factorization scheme.
The same kind of analysis, but for weak boson-gluon fusion processes, is
clearly
more involved due to the presence of different mass scales in the same process
and must be carefully carried out.

In recent years, several works have been presented in connection with spin
dependent weak structure functions \cite{lampe,vogelweb,ravin}. Among them,
those adressing to ${\cal O}(\alpha_{s})$
corrections, only deal with gluon initiated corrections, and, in most cases,
without worring about the proper substraction of soft contributions.
In the present paper we calculate the complete next to leading logarithmic
-quark and gluon initiated- corrections to the spin dependent weak deep
inelastic structure functions, using dimensional regularization \cite{bollini}
and treating $\gamma_{5}$ and $\epsilon_{\mu\nu\rho\sigma}$
according to the original proposal by t'Hooft and Veltman and systematized by
Breitenlohner and Maison \cite{hvbm}. This method, hereafter referenced as
HVBM, has
all the advantages of dimensional regularization, can be implemented at
higher orders of perturbation theory and is free from inconsistencies. In
addition, it has been
shown that soft contributions to different processes involving massless quarks
can clearly be identified and substracted in this scheme \cite{weber}. The
presence of non
zero masses, eventually large compared to the factorization scale $\mu_{fact}$,
changes this
situation requiring a different factorization prescription. This problem is
also
addressed.

Finally, we discuss the phenomenological consequences of the alluded
corrections in
the structure functions, discriminating contributions
from different origins and using parton distributions, defined with the above
mentioned factorization prescription, which fit the available electromagnetic
data.
\\

\noindent {\large \bf Structure functions and parton distributions}\\

In this section we begin establishing the definition of the weak structure
functions
that will be used throughout this paper and also our notation for parton
distributions beyond leading order.

The spin dependent component of the hadronic tensor, in the case of a
longitudinally polarized target, can be written as
\begin{eqnarray}
W^{\mu\nu} & \equiv & -i \epsilon^{\mu\nu\rho\sigma} \frac{q_{\rho}P_{\sigma}}
{P\cdot q}g_{1} + (-g^{\mu\nu}+\frac{q^{\mu}q^{\nu}}{q^{2}})g_{3} \nonumber \\
& & +\frac{1}{P\cdot q}(P^{\mu}-\frac{P\cdot q}{q^{2}}q^{\mu})
(P^{\nu}-\frac{P\cdot q}{q^{2}}q^{\nu})g_{4}
\end{eqnarray}
where $q$ and $P$ are the four-momentum of the  exchanged virtual boson and
the nucleon target respectively. With this definition, the structure functions
have, at lowest order, the following form in terms of parton distributions
\begin{equation}
g_{j}^{B}(x)=\sum_{i}C_{ij}^{B}(x)\, \Delta q_{i}(x)
\end{equation}
where $x$ is the usual Bjorken variable i.e.
\begin{equation}
x=\frac{-q^{2}}{2P\cdot q}
\end{equation}
$\Delta q_{i}$ are the spin dependent quark distributions, and the
coefficients $C_{ij}^{B}$ are given in tables 1 and 2. The index $i$ runs over
the
quark flavours and $B$ indicates the boson  exchanged in the process
under consideration.

The next to leading logarithmic corrections arise, at order
$\alpha_{s}$, from the diagrams in figures 1, 2 and 3.
The evaluation of the diagram in figure 1a corresponds to the lowest order
expressions for the structure functions (equation 2). Diagrams 1b and 1c give
no
contribution for massless quarks in the Landau gauge. The interference
between 1a and 1d, and also the diagrams of figure 2, produce both infrared
and collinear contributions. While the infrared contributions cancel each
other, the divergencies of collinear origin remain and have to be factorized
in the definition of parton distributions. The diagrams of figure 3 also
have collinear divergencies  that have  to be removed in the same way.
In order to deal with the occuring divergencies we use dimensional
regularization.
In the diagrams considered above, $\gamma_{5}$ matrices are present not only
due to the helicity proyectors, but also because of the weak interaction
vertex. As we have mentioned before, we have chosen to deal with this object
following the HVBM proposal. Matrix elements can be straightforwardly
calculated
with the program {\sc Tracer} \cite{tracer}, which masters most of the
intricacies of the
method.

By means of the usual proyector technique, we obtain the $\alpha_{s}$
contributions to the structure functions from the phase space integrated
matrix elements (see appendix A for calculational details)
\begin{eqnarray}
g_{j}^{B}(x)&=&\sum_{i}  \int_{x}^{1} \frac{dz}{z} C_{ij}^{B}(z) \left \{
\left [ \delta (1-z)
+\frac{\alpha_{s}}{2\pi} \Delta P_{qq}(z)(\ln\frac{Q^{2}}{\mu^{2}}
-\frac{1}{\hat{\epsilon}})+\frac{\alpha_{s}}{2\pi} \Delta f^{q}_{j}(z)\right ]
\Delta q_{i}^{0}(\frac{x}{z})\right .\nonumber \\
&&\left . +\left [\frac{\alpha_{s}}{2\pi} \Delta
P_{qg}(z)(\ln\frac{Q^{2}}{\mu^{2}}
-\frac{1}{\hat{\epsilon}})+\frac{\alpha_{s}}{2\pi} \Delta f^{g}_{j}(z)\right ]
\Delta g^{0}(\frac{x}{z})\right \}
\end{eqnarray}
The sum in this equation  runs over the flavours of those quarks and antiquarks
that participate
in the vertex. $\Delta P_{qq}$ and $\Delta P_{qg}$ are the usual
Altarelli-Parisi
evolution kernels \cite{ap}. $\mu$ is the scale introduced by dimensional
regularization and
\begin{equation}
\frac{1}{\hat{\epsilon}}=\frac{1}{\epsilon} +\ln 4\pi
-\gamma_{E}
\end{equation}
where $\epsilon$ is defined through $d=4-2\epsilon$, being $d$ the space-time
dimension. $\Delta q^{0}_{i}$ and $\Delta g^{0}$ are the bare densities of
quarks and gluons respectively.
The term proportional to $\delta (1-z)$ is the lowest order contribution
arising from diagram 1a, while $\Delta f_{q}$ and $\Delta f_{g}$ are the
quark and gluon initiated finite non logarithmic corrections.

Defining the scale dependent next to leading logarithmic quark distributions
as
\begin{eqnarray}
\Delta q_{i}(x,Q^{2})&=&\Delta q_{i}^{0}(x)+\frac{\alpha_{s}}{2\pi}
\int_{x}^{1} \frac{dz}{z} \left [  \Delta P_{qq}(z)(\ln\frac{Q^{2}}{\mu^{2}}
-\frac{1}{\hat{\epsilon}})+ \widetilde{\Delta f^{q}_{i}}(z) \right ]
\Delta q_{i}^{0}(\frac{x}{z}) \nonumber \\
&&+\frac{\alpha_{s}}{2\pi}
\int_{x}^{1} \frac{dz}{z} \left [ \Delta P_{qg}(z)
(\ln\frac{Q^{2}}{\mu^{2}}-\frac{1}{\hat{\epsilon}})
+ \widetilde{\Delta f^{g}_{i}}(z) \right ] \Delta g^{0}(\frac{x}{z})
\end{eqnarray}
the structure functions are then given by
\begin{eqnarray}
g_{j}^{B}(x)&=&\sum_{i} C_{ij}^{B}(x)\Delta q_{i}(x,Q^{2})+
\frac{\alpha_{s}}{2\pi}\sum_{i}  \int_{x}^{1} \frac{dz}{z} C_{ij}^{B}(z)
\left [\Delta f^{q}_{j}(z)-\widetilde{\Delta f^{q}_{i}}(z) \right ]
\Delta q_{i}(\frac{x}{z},Q^{2}) \nonumber \\
&& +\frac{\alpha_{s}}{2\pi}\sum_{i} \int_{x}^{1} \frac{dz}{z}  C_{ij}^{B}(z)
\left [\Delta f^{g}_{j}(z)-\widetilde{\Delta f^{g}_{i}}(z) \right ]
\Delta g(\frac{x}{z},Q^{2})
\end{eqnarray}

This definition factorizes the poles accounting for collinear singularities
and the logarithmic terms associated with the Altarelli-Parisi evolution in
the quark distributions. The terms $\widetilde{\Delta f_{i}^{q}}$ and
$\widetilde{\Delta f_{i}^{g}}$, are designed to absorbe eventual soft
contributions from the non logarithmic corrections. Notice that in the
usual $\overline{MS}$ \cite{ms} scheme this substraction terms are not
included.

Modern parton distributions are usually defined in what is called a variable
(scale dependent) flavour number scheme \cite{collins}. In schemes of this
kind, quarks
whose masses are smaller than the typical energy scale $\mu_{phys}$ are
considered  massless in the evolution.
For those quarks with masses larger than the scale, there are no
associated parton distributions. In this paper we adopt this kind of scheme,
however we keep in mind the absolute mass hierarchy for factorization
purposes.\\

\noindent{\large \bf ${\cal O}(\alpha_{s})$ corrections}\\

In this section we calculate the soft and hard contributions corresponding to
the different diagrams. We begin with those related to the box diagrams of
figure 3. In reference \cite{vogel} this kind of diagram, but for an
electromagnetic
interaction, was evaluated using dimensional regularization within
the HVBM method obtaining
\begin{equation}
g_{1\,(parton)}^{\gamma}=e^{2}_{i} \frac{\alpha_{s}}{2\pi}\Delta P_{qg}
(\ln\frac{Q^{2}}{\mu^{2}}-\frac{1}{\hat{\epsilon}})+e^{2}_{i}\frac{\alpha_{s}}
{2\pi} \frac{1}{2}\left [ (2z-1)(\ln\frac{1-z}{z}-1)+2(1-z)\right ]
\end{equation}
The subscript $parton$ means that in order to obtain the structure function,
the expression must be convoluted with the appropriated bare parton
distribution. Doing this and comparing with the electromagnetic version
of equation (4), $\Delta f_{g}$ can be identified with
\begin{equation}
\Delta f^{g}_{1}(z)=\frac{1}{2} \left [ (2z-1)(\ln\frac{1-z}{z}-1)+2(1-z)\right
]
\end{equation}

It has been shown \cite{vogel} that the last term in equation (9) has a soft
origin in
the case of massless quarks, and therefore must be factorized. This means
that, for massless quarks, $\widetilde{\Delta f^{g}_{i}}$ must be fixed
accordingly
\begin{equation}
\widetilde{\Delta f^{g}}_{m<\mu_{fact}}=1-z
\end{equation}
However if we are dealing with quarks whose masses are smaller than the
physical scale but are larger than the scale that defines soft and hard
phenomena, they can be considered as massless for the evolution (active
flavour)
but the last term in equation (9) corresponds to a hard contribution and should
not be factorized, i.e.
\begin{equation}
\widetilde{\Delta f^{g}}_{m>\mu_{fact}}=0
\end{equation}
Of course, this implies a non vanishing gluonic contribution to the
first moment of  $g_{1}^{em}$ only for light flavours $(m<\mu_{fact})$
independently of which are active $(m<\mu_{phys})$.

The preceeding discussion defines the gluonic part $\widetilde{\Delta
f^{g}_{i}}$
of our factorization prescription. $\widetilde{\Delta f^{q}_{i}}$ will be
fixed after we
evaluate the quark initiated processes. In the following we calculate the
analogue of equation  (9) for the weak structure functions and verify that
our prescription factorizes soft contributions.

For $g_{1}$, the finite non logarithmic contribution $\Delta f_{g}$ is
identical for both the electromagnetic and weak structure functions. The
effect of the weak vertex is completely absorbed in the coefficient
$C_{i1}^{B}$.
If the exchanged boson does not produce change in the quark flavour, the
factorization
prescription works in complete analogy to the electromagnetic case. However,
if there is change of flavour and one of the quarks has a mass larger than
the factorization scale, whereas the other does not, then there is a soft
contribution arising from the lighter but not from the heavier. The
factorization
prescription accounts for this as it should.

For $g_{3}$ and $g_{4}$, $\Delta f_{g}$ vanishes. Due to equation (11), the
contribution associated with the heavier flavours via $\widetilde{\Delta
f^{g}}$
also vanishes. When the exchanged boson is a $Z^{0}$, the contribution from
the lighter quarks and antiquarks cancel between each other leading to
a vanishing gluonic contribution to the structure functions. However, for
flavour changing bosons there is no such cancellation. For example,
if we deal with a $W^{+}$ and in a 4-flavour scheme, only $d$,$\overline{u}$,
$s$ and $\overline{c}$ participate, and using equation (7) we have
\begin{eqnarray}
g_{3}^{W^{+}}(x,Q^{2})\mid_{gluonic} & = &
\frac{\alpha_{s}}{2\pi}\int_{x}^{1} \frac{dz}{z}
\sum_{i=d,\overline{u},s} C_{i3}^{W^{+}} \left [-\widetilde{\Delta
f^{g}_{i}}(z) \right ]
\Delta g(\frac{x}{z},Q^{2}) \nonumber \\
&=& \frac{\alpha_{s}}{2\pi}\int_{x}^{1} dz\, \frac{1-z}{z}\,
\Delta g(\frac{x}{z},Q^{2})
\end{eqnarray}
This result agrees, if the limit $ m^{2}/Q^{2} \rightarrow 0$ is taken,
with that of reference \cite{vogelweb}, where the gluonic contribution
to the structure functions was evaluated using quark masses and a transverse
momentum cut-off as regulator.

The quark initiated non logarithmic correction $\Delta f_{q}$ comes from
the processes in figures 1 and 2. The amplitudes associated to diagrams 2a
and 2b are straightforwardly evaluated and integrated in the corresponding
phase space. Calculated in the Landau gauge and using the HVBM method, the
vertex correction amounts to replace
\begin{equation}
\gamma_{\mu}(1-a\gamma_{5}) \leftrightarrow \gamma_{\mu}(1-a\gamma_{5})
\left \{ 1+ \frac{\alpha_{s}}{4\pi}\frac{4}{3}\left
(\frac{4\pi\mu^{2}}{Q^{2}}\right)^{\epsilon}
\frac{\Gamma(1+\epsilon)\Gamma^{2}(1-\epsilon)}{\Gamma(1-2\epsilon)}
\left [ \frac{-2}{\epsilon^{2}}-\frac{2}{\epsilon}-8 \right ] \right \}
\end{equation}
As in this gauge the quark self energy vanishes at order $\alpha_{s}$ for
massless quarks, there is no quark wave function  renormalization, all the
effect of the interference between diagrams 1a and 1d is the replacement of
the delta function in equation (4) by the delta times the factor between
brackets
in equation (13). Adding this contribution to those of diagrams 2a and 2b,
and isolating the logarithmic and pole terms, we identify the $\Delta f_{q}$
term.

For $g_{1}^{B}$ the non logarithmic correction results to be
\begin{eqnarray}
\Delta f^{q}_{1}(z)&=&\frac{4}{3} \left [ (1+z^{2})\left ( \frac{\ln(1-z)}
{1-z}\right ) _{+}
-\frac{3}{2}\left(\frac{1}{1-z}\right)_{+}-\frac{1+z^{2}}{1-z}\ln z \right
.\nonumber \\
& & \left .+2+z- (\frac{9}{2}+\frac{1}{3}
\pi^{2})\delta (1-z) \right ]-\frac{16}{3}(1-z)
\end{eqnarray}
The last term can be traced back to have a collinear origin for massless
quarks,
so this defines the correspondent tilde-term
\begin{equation}
\widetilde{\Delta f^{q}}_{m<\mu_{fact}}=-\frac{16}{3}(1-z)
\end{equation}
Analogously, we fix for heavy quarks
\begin{equation}
\widetilde{\Delta f^{q}}_{m>\mu_{fact}}=0
\end{equation}
Equations (15) and (16) complete the definition of the factorization
prescription
we adopt. This prescription agrees with the one used in reference \cite{weber}
for massless quarks but is clearly different for massive quarks.

Finally, for $g_{3}$ and $g_{4}$ the corrections are given by
\begin{equation}
\Delta f^{q}_{4}(z)=\Delta f^{q}_{1}(z)+\frac{4}{3}(1+z)
\end{equation}
and
\begin{equation}
\Delta f^{q}_{3}(z)=\Delta f^{q}_{1}(z)+\frac{4}{3}(1-z)
\end{equation}
The next to leading order quark initiated corrections to the structure
functions
$g_{3}$ and $g_{4}$ ($\Delta f^{q}_{3,4} - \widetilde{\Delta f^{q}_{i}}$)
are identical to those obtained for the unpolarized structure functions
$F_{1}$ and $F_{2}$, due to the same tensorial structure at partonic level
\cite{aem}.

In a scheme where the regularization is performed keeping explicitly the masses
of the quarks and introducing a cut off in transverse momentum, the hard
contributions
are obtained taking into account in each process the mass hierarchy.
However, with the factorization prescription we propose, the
non logarithmic terms $\Delta f^{q}$ and $\Delta f^{g}$ are not dependent
on the kind of boson exchanged or the relation between the quark masses and
the factorization scale. This kind of dependence is absorbed in the process
independent $\widetilde{\Delta f}$ terms showing clearly the universal
character of factorization.\\

\noindent{\large \bf Numerical Results}\\

In this section we analyse the relevance of the correction terms we have just
calculated.
All the analysis on ${\cal O}(\alpha_{s})$
corrections to the spin dependent weak structure functions
available in the literature deal only with
the gluon initiated corrections. Even more, most studies on spin dependent
electromagnetic structure functions and parton distributions disregard the
quark initiated corrections either not including them in equation (7)
($\Delta f_{q}=0$) or approximating them by their effect in the moment of
the structure function ($\Delta f_{q}=-\frac{\alpha_s} {\pi}\,\delta (1-z)\,
$).
As the contribution to the moment is of order $\frac{\alpha_{s}}{\pi}$,
it is usually assumed that these corrections are small, however as the $\Delta
f_{q}$
are non trivial functions, nothing guarantees that the corrections are not
comparable to the others for some $x$ interval. Of course, one can always
define a factorization prescription in such a way that some correction terms
are absorbed in the parton distributions (at least for certain structure
functions)
provided the prescription is implemented consistently in other processes.
This, however changes radically the interpretation of parton distributions.
It is important then,
to have an estimate of the relative weight of the corrections within the
scheme employed and particularly in one that factorizes soft contributions
in each process.

We begin comparing numerically the quark and gluon initiated
corrections to the structure functions (second and third terms in equation
(7), respectively), between each other and with the tree level part
(first term of the same equation).
In order to make the estimates, we need a set of spin dependent parton
distributions
defined at order $\alpha_{s}$ and within the factorization scheme proposed.
We build the set
taking for the bare densities the functional forms
suggested by the spin dilution model \cite{dist} and fix the free parameters
of it in such a way that the available data on spin dependent electromagnetic
structure functions is reproduced.
This procedure reduces considerably the number of parameters to be fixed and
includes in the analysis other theoretical and experimental ingredients
which are detailed in reference \cite{dist}.

As the next to leading order evolution kernels for the polarized case
haven't been calculated yet, in order to evaluate data at different values
of $Q^{2}$ one can either use leading order evolution kernels or consider
observables with a moderate dependence as the asymmetries at an average
$Q^{2}$ value of $10\,GeV^{2}$. It has been shown \cite{anr} that the effect
of the $Q^{2}$ evolution in the asymmetries is small when compared to the
ambiguities associated
with the fitting procedure and the experimental errors. Figures (4) and (5)
show the agreement between the electromagnetic data and the asymmetries
calculated
with the set.

In Figures (6), (7) and (8) we show the contributions to different
structure functions $x\,g_{1}^{\gamma}(x)$, $x\,g_{1}^{W^{+}}(x)$ and
$x\,g_{3}^{W^{+}}(x)$, discriminating their different origins (naive,
quark initiated correction, gluon initiated correction). The figures clearly
show that even though the moment of the quark initiated corrections  are
smaller than those of the gluons, which have it main contribution in the
very small $x$ region, the quark initiated corrections are greater
($x\,g_{1}^{\gamma}(x)$,
$x\,g_{3}^{W^{+}}(x)$) or comparable ($x\,g_{1}^{W^{+}}(x)$) to the gluonic
ones in most of the $x$ interval. Compared to the naive contribution, the
quark initiated correction can be as large as $20\%$ of the former, while
the $\frac{\alpha_{s}}{\pi}$ approximation ammount to $7\%$. Notice also,
that there is a non vanishing gluon contribution to $x\,g_{3}^{W^{+}}(x)$.
In this structure function the gluon contributions associated with quarks
$\overline{u}$ and $d$ in the box diagram, cancel each other while there
is no such cancellation between $s$ and $\overline{c}$ due to the different
factorization properties they have (equation 11). \\

\noindent{\large \bf Conclusions}\\

We have computed the complete ${\cal O}(\alpha_{s})$ corrections to the
spin dependent deep inelastic scattering weak structure functions introducing
a factorization prescription that safely removes soft contributions taking
into account the problem of mixing different mass scales in a same process.
Using a set of spin dependent parton distributions that reproduces the
electromagnetic
data, we have shown that the quark initiated corrections are larger or
comparable to those
gluon initiated even though their moments are not.\\

\noindent{\large \bf Acknowledgements}\\

We thank L.N.Epele, H.Fanchiotti and C.A.Garc\'{\i}a Canal for helpful
discussions and carefully reading the manuscript.\\

\noindent{\large \bf Appendix A}\\

This appendix contains some calculational details related to the
definitions of structure functions at partonic level.

In addition to the hadronic tensor (eq.1), one can also define a partonic one
which has a similar structure, but is written in terms of partonic structure
functions.
\begin{eqnarray}
w^{\mu\nu} & \equiv & -i \epsilon^{\mu\nu\rho\sigma} \frac{q_{\rho}p_{\sigma}}
{p\cdot q}g_{1}^{partonic} +
(-g^{\mu\nu}+\frac{q^{\mu}q^{\nu}}{q^{2}})g_{3}^{partonic} \nonumber \\
& & +\frac{1}{p\cdot q}(p^{\mu}-\frac{p\cdot q}{q^{2}}q^{\mu})
(p^{\nu}-\frac{p\cdot q}{q^{2}}q^{\nu})g_{4}^{partonic}
\end{eqnarray}
The phase-space integrated matrix elements of either the quark or the gluon
initiated diagrams of figs.1, 2 and 3 give the quark and gluon components of
the partonic tensor, respectively.
The partonic structure functions must be convoluted with the appropriated
bare parton distributions in order to obtain the usual hadronic ones.

In order to isolate each partonic structure function it is customary to
define the following proyectors
\begin{eqnarray}
P_1^{\mu\nu}&=& i \epsilon^{\mu\nu\rho\sigma} \frac{q_{\rho}p_{\sigma}}{2
p\cdot q}\nonumber\\
P_3^{\mu\nu}&=& \frac{1}{2(1-\epsilon)}\left[ -g^{\mu\nu}+\frac{4
z^2}{Q^2}p^{\mu}p^{\nu} \right]\\
P_4^{\mu\nu}&=& \frac{z}{(1-\epsilon)}\left[ - g^{\mu\nu}+\frac{4
z^2(3-2\epsilon)}
{Q^2}p^{\mu}p^{\nu} \right]\nonumber
\end{eqnarray}
where
\begin{equation}
z=\frac{Q^2}{2 p\cdot q}.
\end{equation}
This implies
\begin{equation}
g_i^{partonic}=\frac{1}{4 \pi} \int d\Gamma\,\,\Delta\left| M\right|
^2_{\mu\nu} P_i^{\mu\nu}
\end{equation}
where the spin dependent amplitude  $\Delta\left| M\right| ^2_{\mu\nu}=
\frac{1}{2}\left[\left| M_+\right| ^2_{\mu\nu}-\left| M_-\right|
^2_{\mu\nu}\right]$
is defined in terms of those for partons whose polarization is parallel($+$)
or antiparallel($-$) to that of the target and is normalized so that the
tree-level diagram fig.1a gives a $\delta (1-z)$ .
The n-dimensional phase-space $d\Gamma$ within the HVBM method is given by

\begin{equation}
\int d\Gamma=\frac{1}{8\pi} \frac{(4\pi)^\epsilon}{\Gamma(-\epsilon)}
\int_0^1 dy \int_0^{sy(1-y)} d\hat{k}^2\, \hat{k}^{-2(1+\epsilon)}
\end{equation}
for two outgoing particles. $\hat{k}_\mu$ is the n-4 dimensional component
of the momentum of one of the outgoing partons \cite{vogel}.\\

\noindent{\large \bf Appendix B}\\

The bare parton distributions inspired in the spin dilution model and used
in this paper can be effectively parametrized as
\begin{equation}
x\,\Delta q(x)=A_{q}\,x^{B_{q}}\,(1-x)^{C_{q}}\,(1+D_{q}\,x+E_{q}\,x^{2})
\end{equation}
where the parameters are given in Table (3). Notice that these parameters
are not the free parameters of the modified spin dilution model to be
adjusted!. The former include information about the unpolarized parton
distributions.

\pagebreak

\pagebreak
\noindent{\large \bf Figure Captions}
\\

\begin{enumerate}
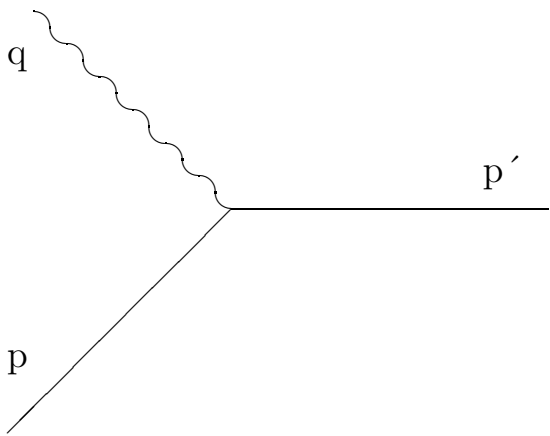
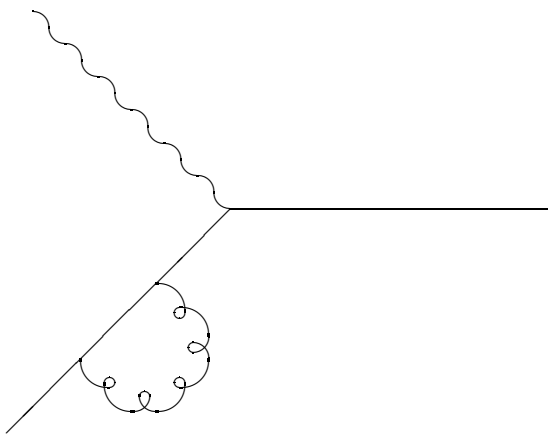
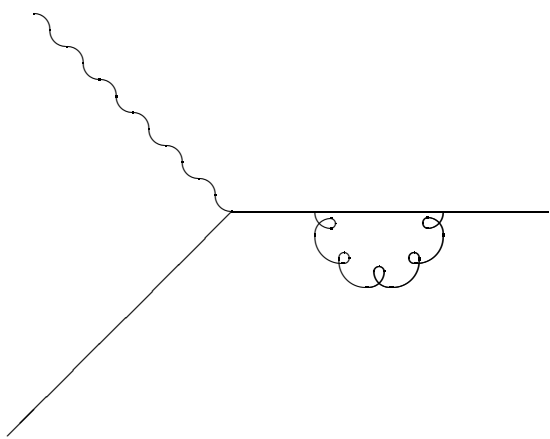
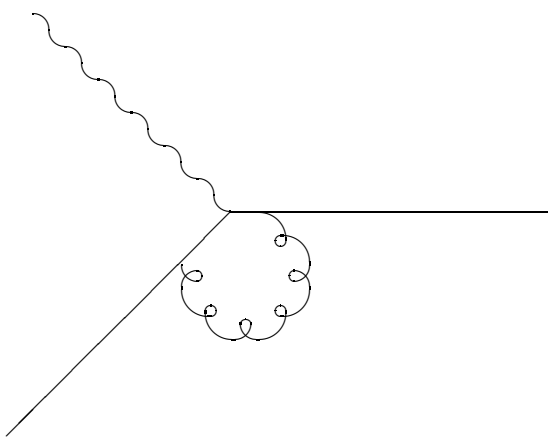
\item[Figure 1 ] a) Lowest order graph for DIS; b),c) and d) virtual
gluon correction graphs to a).
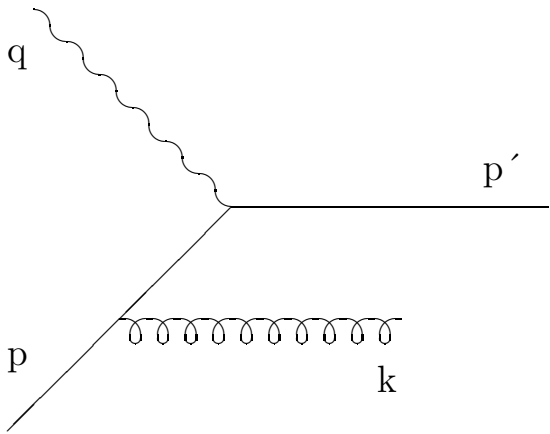
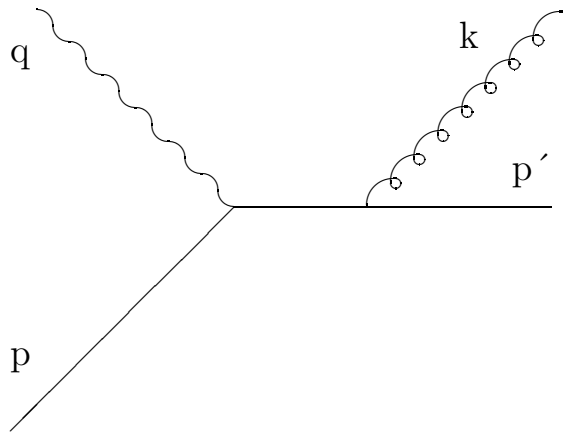
\item[Figure 2 ] Real gluon emission corrections to 1a).
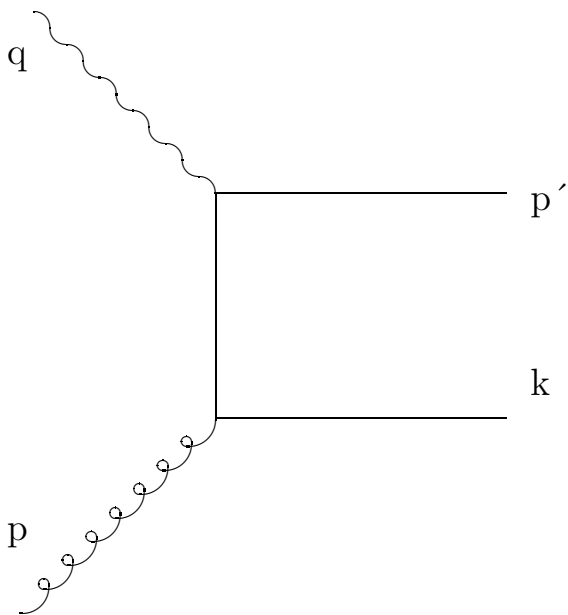
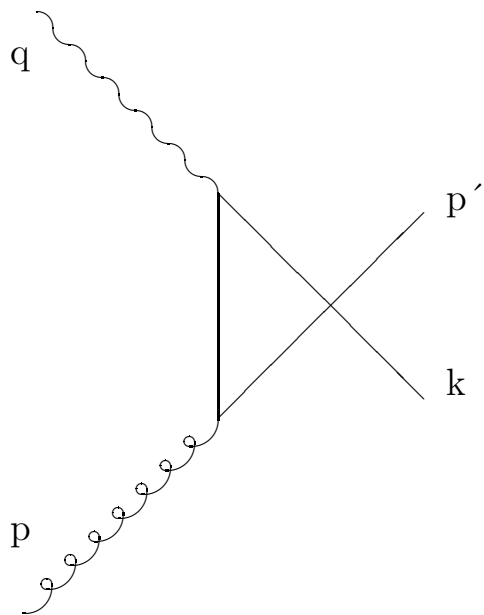
\item[Figure 3 ] Gluon contribution to DIS at order $\alpha_{s}$
\item[Figure 4 ]  The spin-dependent proton asymmetry given by the fit
compared to SMC, E-143 and earlier EMC data.
\item[Figure 5 ] The spin-dependent neutron asymmetry given by the fit
compared to E-142 data and the combined proton and deuteron SMC data.
\item[Figure 6 ] The the naive contribution and the quark and gluon initiated
corrections to the spin-dependent structure function $x\,g_{1}^{\gamma}(x)$
\item[Figure 7 ] The same as figure 6 for $x\,g_{1}^{W^{+}}(x)$ .
\item[Figure 8 ] The same as figure 6 for $x\,g_{3}^{W^{+}}(x)$.
\end{enumerate}
\vspace*{10mm}
\noindent{\large \bf Table Captions}\\

\begin{enumerate}
\item[Table 1 ]  Values for the coefficients $C_{ij}^{B}$ defined in equation
(2).
\item[Table 2 ]  Axial and  vector couplings for the $Z^{0}$ exchange.
\item[Table 3 ]  Parameters of the spin dependent parton distributions.
\end{enumerate}

\pagebreak
\vspace*{5mm}
\vspace*{5mm}
\begin{center}
\begin{tabular}{|c|c|c|c|}\hline
      & $B=\gamma$            & $B=W^{+}$ & $B=Z^{0}$ \\ \hline
$j=1$ & $e_{i}^{2}/2$ &  $1$      & $(C_{Vi}^{2}+C_{Ai}^{2})/2$ \\
$j=3 \,\,\, \left \{ \begin{array}{l} i=q \\ i=\overline{q}  \end{array}
\right. $
& $\left . \begin{array}{c}  0\\  0 \end{array} \right.$
& $\left . \begin{array}{c}  -1\\  1 \end{array} \right.$
&$\left. \begin{array}{c}  -C_{Vi}C_{Ai}\\   C_{Vi}C_{Ai} \end{array} \right.$
\\
$j=4\,\,\, \left \{ \begin{array}{l} i=q \\ i=\overline{q}  \end{array} \right.
$
& $\left . \begin{array}{c}  0\\  0 \end{array} \right.$
&$\left . \begin{array}{c}  -2z\\   2z \end{array} \right.$
&$\left . \begin{array}{c}  -2z\, C_{Vi}C_{Ai}\\   2z\,C_{Vi}C_{Ai} \end{array}
\right.$

\\  \hline
\end{tabular}
\vspace*{5mm}

{\bf Table 1:} Values for the coefficients $C_{ij}^{B}$ defined in equation
(2).
\vspace*{5mm}
\vspace*{5mm}

\begin{tabular}{|c|c|c|}\hline
& $i=u,c,t$ & $i=d,s,b$ \\ \hline
$C_{Vi}$ & $ \frac{1}{2}-\frac{4}{3}sin^{2}\theta_{W}$
& $ -\frac{1}{2}+\frac{2}{3}sin^{2}\theta_{W}$ \\
$C_{Ai}$ & $ \frac{1}{2}$ & $-\frac{1}{2}$ \\ \hline
\end{tabular}
\vspace*{5mm}

{\bf Table 2:} Axial and  vector couplings for the $Z^{0}$ exchange.
\vspace*{5mm}
\vspace*{5mm}

\begin{tabular}{|c|c|c|c|c|c|}\hline
              & $A$         & $B$    & $C$     & $D$    & $E$ \\ \hline
$u_{v}$       & 1.052       & 0.704  & 4.085   & 2.467  & 17.34 \\
$d_{v}$       & -1.066      & 0.7685 & 2.179   & -1.645 & 0.4527 \\
$\overline{u}$& 1.956       & 1.504  & 4.676   & -3.391 & 2.981 \\
$\overline{d}$& 1.035       & 1.734  & 2.905   & -3.051 & 2.324 \\
$\overline{s}$& 0.0932      & 1.594  & 8.694   & -2.149 & 32.709 \\
$\overline{c}$& -2.171      & 57.27  & 173.6   & -20.36 & 2.279 \\
 $g$          & 3.828       & 0.5243 & 2.287   & -2.747 & 1.953 \\ \hline
\end{tabular}
\vspace*{5mm}

{\bf Table 3:} Parameters of the spin dependent parton distributions.
\end{center}
\vspace*{5mm}

\newpage
\pagestyle{empty}
\input FEYNMAN
\textheight 400pt \textwidth 200pt
\begin{picture}(10000,18000)
\drawline\photon[\SE\REG](-2000,18000)[12]
\put(-3000,16000){\large  q}
\drawline\fermion[\E\REG](\photonbackx,\photonbacky)[12000]
\put(-3000,4500){\large p}
\drawline\fermion[\SW\REG](\photonbackx,\photonbacky)[12000]
\put(15000,11500){\large p\'}
\put(4000,-1000){\large Figure 1.a}
\drawline\photon[\SE\REG](21000,18000)[12]
\drawline\fermion[\E\REG](\photonbackx,\photonbacky)[12000]
\drawline\fermion[\SW\REG](\photonbackx,\photonbacky)[4000]
\drawloop\gluon[\SE\FLIPPEDFLAT](\fermionbackx,\fermionbacky)
\drawline\fermion[\SW\REG](\fermionbackx,\fermionbacky)[8000]
\put(29000,-1000){\large Figure 1.b}

\drawline\photon[\SE\REG](-2000,-9000)[12]
\drawline\fermion[\E\REG](\photonbackx,\photonbacky)[8000]
\drawloop\gluon[\S\FLIPPEDFLAT](\fermionbackx,\fermionbacky)
\drawloop\gluon[\S\FLIPPEDCURLY](\fermionbackx,\fermionbacky)
\drawline\fermion[\E\REG](\fermionbackx,\fermionbacky)[4000]
\drawline\fermion[\SW\REG](\photonbackx,\photonbacky)[12000]
\put(4000,-29000){\large Figure 1.c}

\drawline\photon[\SE\REG](21000,-9000)[12]
\drawline\fermion[\E\REG](\photonbackx,\photonbacky)[1000]
\drawloop\gluon[\SE\CENTRAL](\fermionbackx,\fermionbacky)
\drawline\fermion[\E\REG](\fermionbackx,\fermionbacky)[11000]
\drawline\fermion[\SW\REG](\photonbackx,\photonbacky)[12000]
\put(29000,-29000){\large Figure 1.d}

\end{picture}
\newpage
\begin{picture}(10000,18000)
\drawline\photon[\SE\REG](-2000,18000)[12]
\put(-3000,16000){\large  q}
\drawline\fermion[\E\REG](\photonbackx,\photonbacky)[12000]
\drawline\fermion[\SW\REG](\photonbackx,\photonbacky)[6000]
\drawline\gluon[\E\REG](\fermionbackx,\fermionbacky)[10]
\drawline\fermion[\SW\REG](\fermionbackx,\fermionbacky)[6000]
\put(-3000,4500){\large p}
\put(15000,11500){\large p\'}
\put(4000,-1000){\large Figure 2.a}
\put(11000,3500){\large k}

\drawline\photon[\SE\REG](21000,18000)[12]
\drawline\fermion[\SW\REG](\photonbackx,\photonbacky)[12000]
\drawline\fermion[\E\REG](\photonbackx,\photonbacky)[5000]
\drawline\gluon[\NE\REG](\fermionbackx,\fermionbacky)[7]
\drawline\fermion[\E\REG](\fermionbackx,\fermionbacky)[7000]
\put(20000,16000){\large  q}
\put(29000,-1000){\large Figure 2.b}
\put(20000,4500){\large p}
\put(39000,11500){\large p\'}
\put(37000,16500){\large k}

\drawline\photon[\SE\REG](-2000,-9000)[11]
\drawline\fermion[\S\REG](\photonbackx,\photonbacky)[8500]
\drawline\gluon[\SW\REG](\fermionbackx,\fermionbacky)[7]
\drawline\fermion[\E\REG](\photonbackx,\photonbacky)[11000]
\drawline\fermion[\E\REG](\gluonfrontx,\gluonfronty)[11000]
\put(4000,-35000){\large Figure 3.a}
\put(-3000,-29000){\large  p}
\put(-3000,-11000){\large  q}
\put(16800,-16500){\large  p\'}
\put(16800,-23500){\large  k}

\drawline\photon[\SE\REG](21000,-9000)[11]
\drawline\fermion[\S\REG](\photonbackx,\photonbacky)[8500]
\drawline\gluon[\SW\REG](\fermionbackx,\fermionbacky)[7]
\drawline\fermion[\SE\REG](\photonbackx,\photonbacky)[11000]
\drawline\fermion[\NE\REG](\gluonfrontx,\gluonfronty)[11000]
\put(29000,-35000){\large Figure 3.b}
\put(20000,-29000){\large  p}
\put(20000,-11000){\large  q}
\put(36500,-16500){\large  p\'}
\put(36500,-23500){\large  k}

\end{picture}
\end{document}